\documentclass[10pt,letterpaper]{article}
\usepackage{amsmath}
\usepackage[T1]{fontenc}
\usepackage{graphicx}
\usepackage{cite}
\usepackage[labelformat=simple]{subcaption}
\makeatletter
  \DeclareCaptionLabelFormat{bparen}{\textbf{(\alph{sub\@captype})}} 
  \renewcommand{\eqref}[1]{\textup{Eq.~\tagform@{\ref{#1}}}} 
\makeatother
\usepackage{opex3}
\usepackage{hyperref}
\hypersetup{
   bookmarks=false,
   colorlinks,
   citecolor=blue,
   filecolor=black,
   linkcolor=black,
   urlcolor=blue
}
\newcommand{\degree}{\ensuremath{^\circ}}

\newcommand{\reffig}{Fig.}
\newcommand{\micron}{\ensuremath{\mu}m~}
\newcommand{\etal}{\textit{et. al.}~}
\newcommand{\e}{\mathrm{e}}
\usepackage[disable]{todonotes} 
\presetkeys{todonotes}{backgroundcolor=blue!20!white, linecolor=blue!50!white}{}
\let \mr=\mathrm

\begin{document}
\title{Novel laser machining of optical fibers for long cavities with low birefringence}

\author{Hiroki Takahashi,$^{1,2,*}$
Jack Morphew,$^{1,3}$
Fedja Oru\v{c}evi\'{c},$^{1,4}$
Atsushi Noguchi,$^{5,6}$
Ezra Kassa$^1$ and
Matthias Keller$^{1}$
}

\address{$^1$Department of Physics and Astronomy, University of Sussex, Brighton, BN1 9QH, UK\\
$^2$PRESTO, Japan Science and Technology Agency (JST), 4-1-8, Honcho
Kawaguchi, Saitama 332-0012 Japan\\
$^3$School of Physics and Astronomy, University of Southampton, Southampton
SO17 1BJ, UK\\
$^4$Midlands Ultracold Atom Research Centre, School of Physics and Astronomy, University of Nottingham, Nottingham NG7 2RD, UK\\
$^5$Graduate school of Engineering Science, Osaka University, 1-3 Machikaneyama, Toyonaka, Osaka, Japan\\
$^6$Current address: Research Center for Advanced Science and Technology (RCAST), The University of Tokyo, Meguro-ku, Tokyo 153-8904, Japan
}

\email{$^*$ht74@sussex.ac.uk}
\begin{abstract}
We present a novel method of machining optical fiber surfaces with a CO${}_2$ laser for use in Fiber-based Fabry-Perot Cavities (FFPCs).
Previously FFPCs were prone to large birefringence and limited to relatively short cavity lengths ($\le$ 200 $\mu$m). These characteristics hinder their use in some applications such as cavity quantum electrodynamics with trapped ions. We optimized the laser machining process to produce large, uniform surface structures. This enables the cavities to achieve high finesse even for long cavity lengths. By rotating the fibers around their axis during the laser machining process the asymmetry resulting from the laser's transverse mode profile is eliminated. Consequently we are able to fabricate fiber mirrors with a high degree of rotational symmetry, leading to remarkably low birefringence.
Through measurements of the cavity finesse over a range of cavity lengths and the polarization dependence of the cavity linewidth, we confirmed the quality of the produced fiber mirrors for use in low-birefringence FFPCs.
\end{abstract}


\bibliographystyle{osajnl}


\section{Introduction}
\label{sec:introduction}

Cavity quantum electrodynamics (cQED) has developed from a research topic of purely fundamental interest to a versatile tool in quantum information processing \cite{Kimble2008}. In order to obtain strong coupling between atomic particles and a cavity, and to create compact systems, several implementations of micro-cavities have been pursued \cite{Vernooy1998a,Armani2003,Hunger2010,pollinger2009ultrahigh,Thompson2013}. Among these fiber-based Fabry-Perot cavities (FFPC) are attracting increasing attention.
By employing CO$_2$ laser machining to form a concave surface on the fiber facet, low roughness and small radii of curvature can be achieved. Cavity finesses in excess of 100,000 have been demonstrated \cite{Brandstatter2013}. Due to their quality and compactness, FFPCs are employed in a wide range of experiments \cite{Colombe2007,Steiner2013,Flowers-Jacobs2012}. 

In previous studies \cite{Hunger2010,Brandstatter2013,Hunger2012} fibers were machined through single-shot laser ablation with a CO$_2$ laser. Due to the imperfect laser beam profile and the highly non-linear dependence of the ablation process on the local laser intensity, the resulting concave surfaces tended to be elliptical.
As a result, the cavity often exhibited birefringence \cite{Brandstatter2013} which hampers their use in applications where superpositions of light polarizations are exploited \cite{Hijlkema2007,Ritter2012}. 
Employing focused single-shot laser ablation usually results in concave fiber surfaces with small radii of curvature and small indentation depths. This in turn limits the lengths of the cavities due to clipping losses at the edge of the indentation. Even though in many applications short cavities are advantageous, in some cases cavities with a length of several hundreds of $\mu$m are required. Among these applications is the coupling of atomic ions to a fiber cavity \cite{Steiner2013}. Due to the strong disturbance of the trapping electric potential  by the presence of the fibers, trapping of ions in the close vicinity of a fiber mirror is challenging and hence relatively long cavity lengths 
are desirable.
 
In this paper we report a novel method to fabricate FFPCs through CO$_2$ laser machining which leads to small birefringence and supports cavity lengths of several hundreds of $\mu$m.  We apply up to 100 laser pulses per fiber each with a large focus between 90 to 120 \micron beam waist. Note that in this paper, beam waist refers to the radius of the Gaussian TEM$_{00}$ mode at which the field amplitude drops to $1/\e$ of its maximum.
While a single laser pulse has only a minor effect on the surface shape, the cumulative effect resulting after applying multiple laser pulses allows for precise tuning of relevant surface parameters such as radius of curvature and the diameter of the indentation by adjusting the number of pulses employed.
Additionally, rotation of the fiber between laser pulses was introduced in order to reduce the ellipticity of the machined surface. The resultant ellipticity of the mirror surfaces are very small, producing cavities with a birefringence below our measurement sensitivity.

We have produced and tested fiber mirrors with a range of radii of curvature from 100 $\mu$m to 700 $\mu$m. All the fiber cavities that we produced show a finesse of 40,000 - 60,000 which is stable over a large range of cavity length.

In Section \ref{sec:laser-mach-fiber}, we describe the optical set-up, the process of laser machining and the characterization of the shape of the fiber end facets. The characterization of the FFPCs is discussed in Section \ref{sec:char-fiber-cavit}.

\section{Laser machining of fiber facets}
\label{sec:laser-mach-fiber}

\subsection{Optical setup}
\label{sec:optical-setup}

\begin{figure}[hbt]
 \centering
 \begin{subfigure}{0.6\textwidth}
  \caption{} \includegraphics[width=\linewidth]{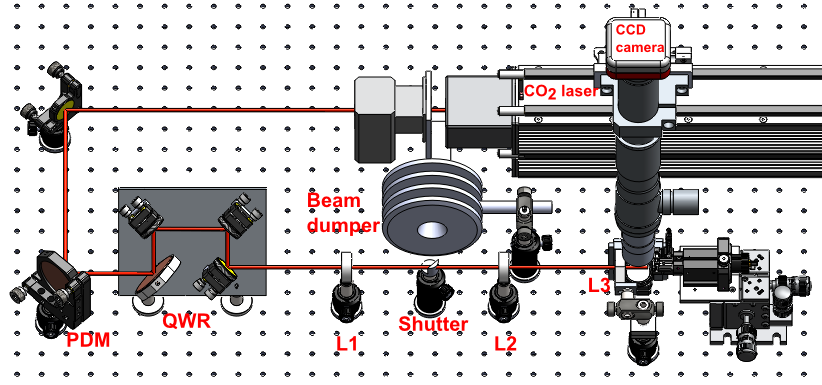}
  \label{fig:lasersetup}
 \end{subfigure}
 \quad
 \begin{subfigure}{0.35\textwidth}
  \caption{}
  \includegraphics[width=\linewidth]{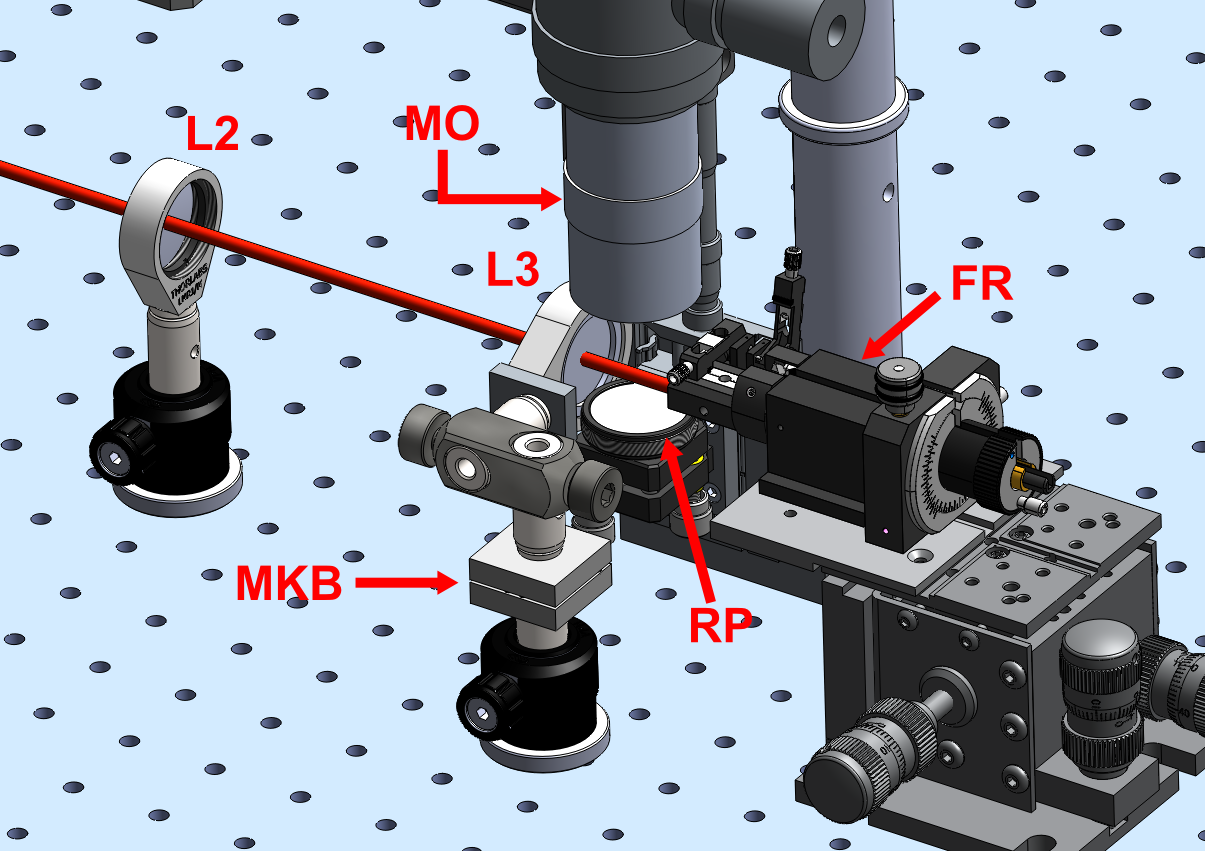}
  \label{fig:lasersetup-zoom}
 \end{subfigure}
 \caption{{\bf \subref{fig:lasersetup}} A schematic of the optical setup for CO$_2$ laser machining. The red line represents the laser trajectory for an open shutter, when closed it is reflected into the beam dumper. The polarization dependent mirror (PDM) reflects vertically polarized light whereas it absorbs horizontally polarized light. The quarter wave retarder (QWR) acts like a quarter wave plate, rotating the incoming vertical polarization into a circular polarization. {\bf \subref{fig:lasersetup-zoom}} A close-up view of the machining setup. MO = microscope objective, RP = reference plate (see Section \ref{sec:surf-analys-interf}), FR = fiber rotator and MKB = magnetically coupled kinematic base.}
\end{figure}

To shape the end facet of an optical fiber by thermal ablation, we employ a CO$_2$ laser (SYNRAD 48-1 water-cooled) in quasi-continuous wave operation \cite{Hunger2012}.
A schematic of the optical setup is shown in \reffig\ref{fig:lasersetup}.
A combination of a polarization dependent mirror (PDM; II-VI Infrared absorbing thin-film reflector) and quarter-wave retarder (QWR; II-VI Infrared Cu reflective phase retarder) work together as an optical isolator to protect the laser. Any back-scattered laser light  after the QWR is absorbed by the PDM. Lens L1 focuses the laser onto a mechanical shutter which controls the duration of a machining pulse. The rise and fall time of the shutter is approximately 1 ms. A second lens, L2, forms a telescopic setup with L1 and is mounted on a linear stage to adjust the expansion of the beam.
A third lens, L3, is placed in the laser path to focus the beam onto the fiber tip. 




This final lens is mounted using a magnetically coupled kinematic base (Thorlabs KB1X1), allowing routine interchangeability with other optics without disturbing the beam alignment. A vertically mounted microscope is employed to analyze the fiber surface. Replacing the magnetically mounted lens with a 45-degree angled mirror enables basic optical inspection, and a beam splitter cube facilitates a more detailed view by white light interferometry, discussed in section \ref{sec:surf-analys-interf}. In this way it is possible to assess the quality of the machined surface {\it in situ} and continue with laser machining until the desired surface parameters are reached.


The fiber is clamped in a v-groove which in turn is part of a fiber rotation stage (Newport 466A-717). The rotation angle is controlled by a stepper motor
to provide reproducible rotation steps without significantly influencing the fiber alignment. In order to align the position of the mounted fiber with the laser beam, the rotation stage is mounted on a 3-axis linear translation stage. Due to imperfection in matching the two axes of the fiber rotor and laser beam, the fiber precesses around the beam by a small amount as it is rotated. This precession can be reduced to a radius of $\sim$10 \micron by careful alignment but it still has a non-negligible impact on the surface symmetry. Therefore we realigned the fibers with the microscope after every rotation. 

\subsection{Laser machining}
\label{sec:laser-shooting}

\begin{figure}[hbt]
 \centering
 \quad
 \begin{subfigure}{0.3\textwidth}
  \caption{}
  \includegraphics[width=\linewidth]{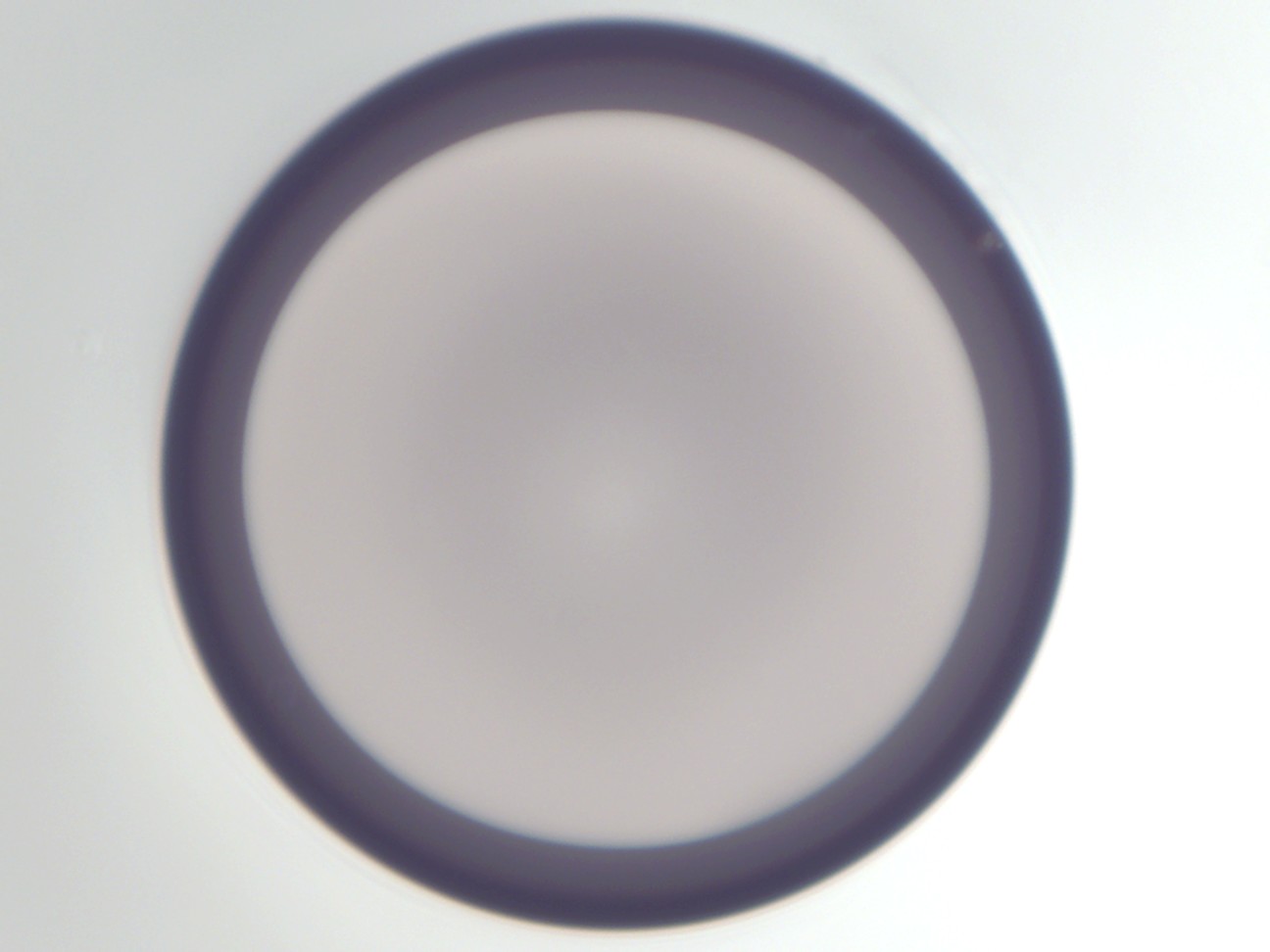}
  \label{fig:rot}
 \end{subfigure}
\hspace{1cm}
 \begin{subfigure}{0.3\textwidth}
  \caption{}
  \includegraphics[width=\linewidth]{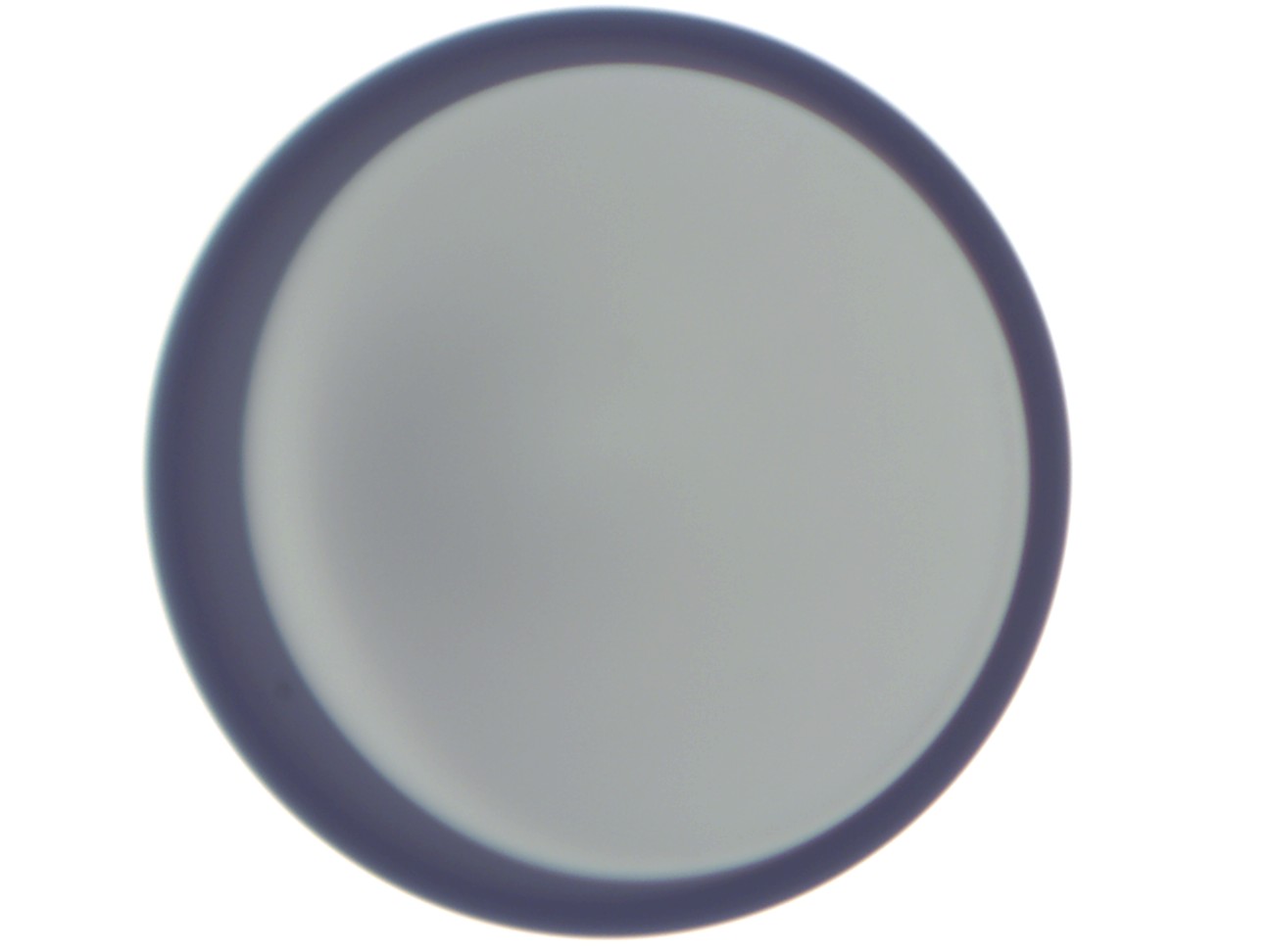}
  \label{fig:nonrot}
 \end{subfigure}

 \caption{Comparison of fiber machining techniques. Optical images of the facets of machined MM-fibers are shown. The edges of the facets are rounded off by the laser shots and they appear to be black rims in the images.
 {\bf\subref{fig:rot}} Produced by 42 laser pulses at 1.43 W, with a beam waist of 93 \micron and duration of 80 ms. The fiber is rotated over uniformly distributed angles.  {\bf\subref{fig:nonrot}} Produced with the same settings as \subref{fig:rot} but without rotation.}
\label{fig:fibercompare}
\end{figure}

To machine a fiber, the laser is pulsed using the mechanical shutter, exposing the fiber tip to the circularly polarized laser for a certain amount of time at each shot. Typical pulses are between 50-80 ms in duration. Laser parameters such as intensity, beam waist and the duration of the pulse all contribute to the thermal ablation process. We have chosen these parameters such that each single pulse has a moderate impact on the fiber surface. This allows us to iteratively refine the surface shape of the fiber. The surface quality and overall symmetry is further enhanced by rotating the fiber by a certain angle (e.g. 45$\degree$) after every few pulses, eliminating effects from the laser beam's asymmetry.
Figure \ref{fig:fibercompare} shows how these techniques have a large impact on the outcome surface structure, with the left image depicting a fiber surface machined with rotation.

We have machined multimode (MM) and single-mode (SM) optical fibers manufactured for infrared wavelengths (Oxford Electronics HPSIR 200CB and IVG fiber Cu800-200). Their core and cladding diameters are 200 and 214 $\mu$m respectively for the MM fibers and 6 and 200 $\mu$m for the SM fibers. All the fibers are coated with a copper layer of 34 $\mu$m (MM) or 25 $\mu$m (SM). In addition to fibers with curved surfaces, we have produced fibers with flat surfaces. In order to clean and smooth the flat surface after cleaving, we apply a few (3-10) pulses with a large beam waist of $\sim$120 \micron.    

\subsection{Surface analysis by interferogram}
\label{sec:surf-analys-interf}

\begin{figure}[h!]
 \centering
 \includegraphics[width=0.45\textwidth]{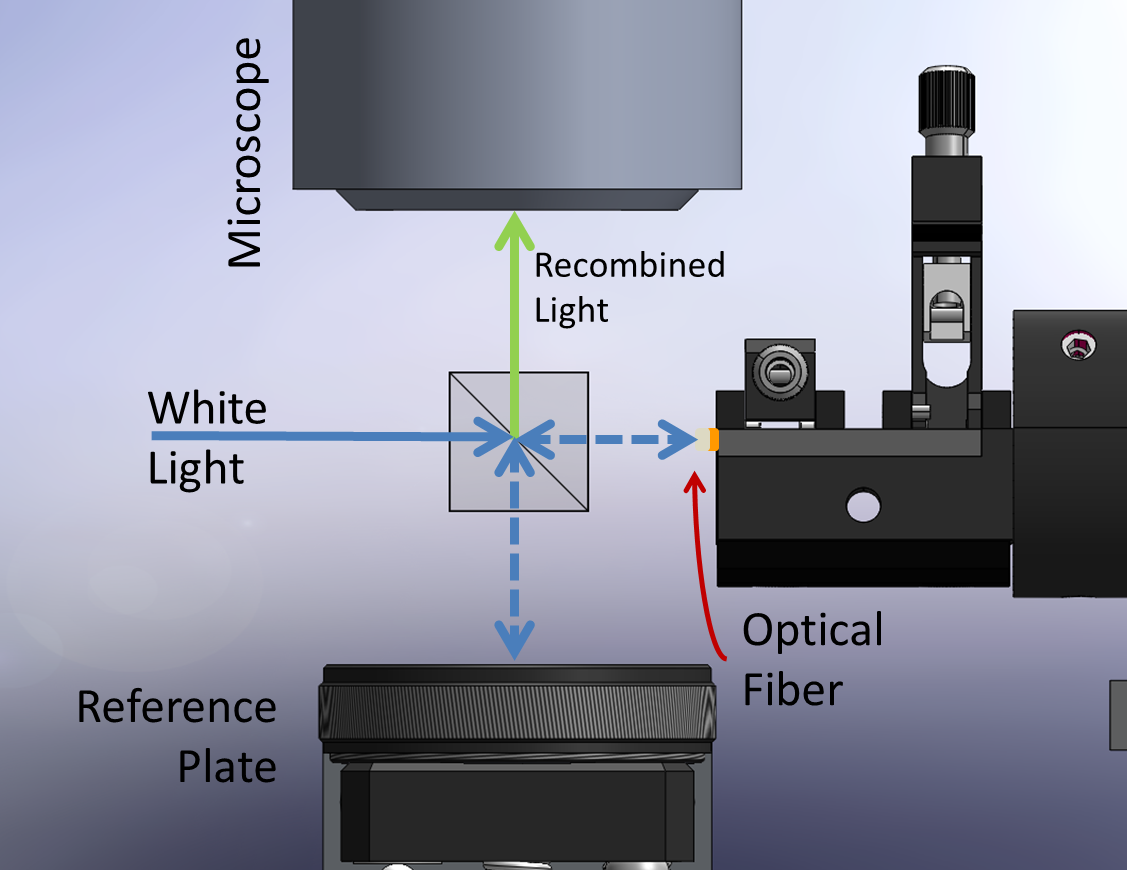}
 \caption{Interferometry setup for the detailed inspection of a machined fiber. The reference plate has a similar reflection coefficient as the fiber tip for a good fringe contrast in white light interferometry. Furthermore it is mounted on a mirror mount in order to align its orientation with respect to the fiber facet.}
 \label{fig:interferencesetup}
\end{figure}

\begin{figure}[hbt]
 \centering
 \begin{subfigure}{0.3\textwidth}
  \caption{}
  \includegraphics[width=\linewidth]{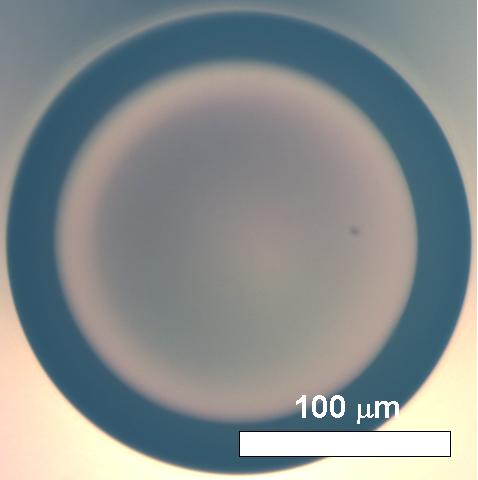}
  \label{fig:f054-opt}
 \end{subfigure}
 \quad
 \begin{subfigure}{0.3\textwidth}
  \caption{}
  \includegraphics[width=\linewidth]{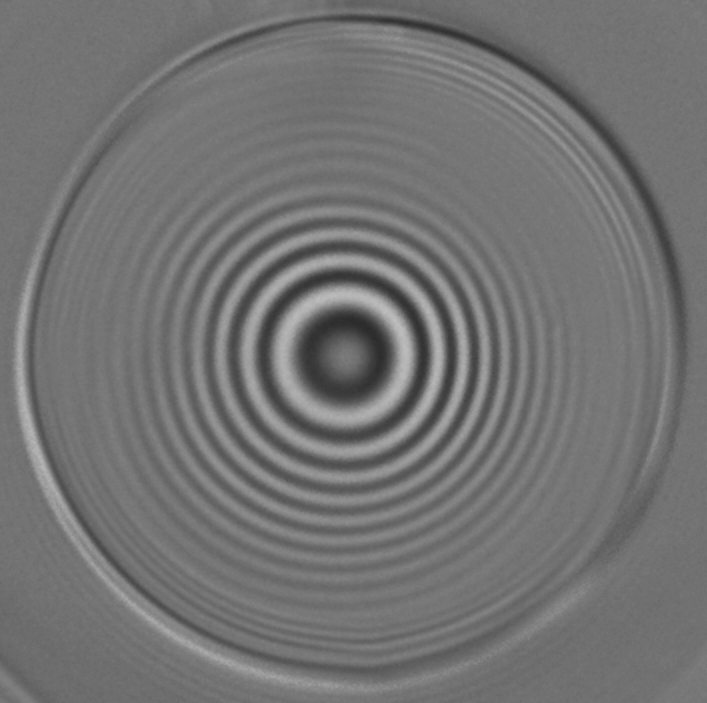}
  \label{fig:f054-int}
 \end{subfigure}
 \\
 \begin{subfigure}{0.5\textwidth}
  \caption{}
  \includegraphics[width=\linewidth]{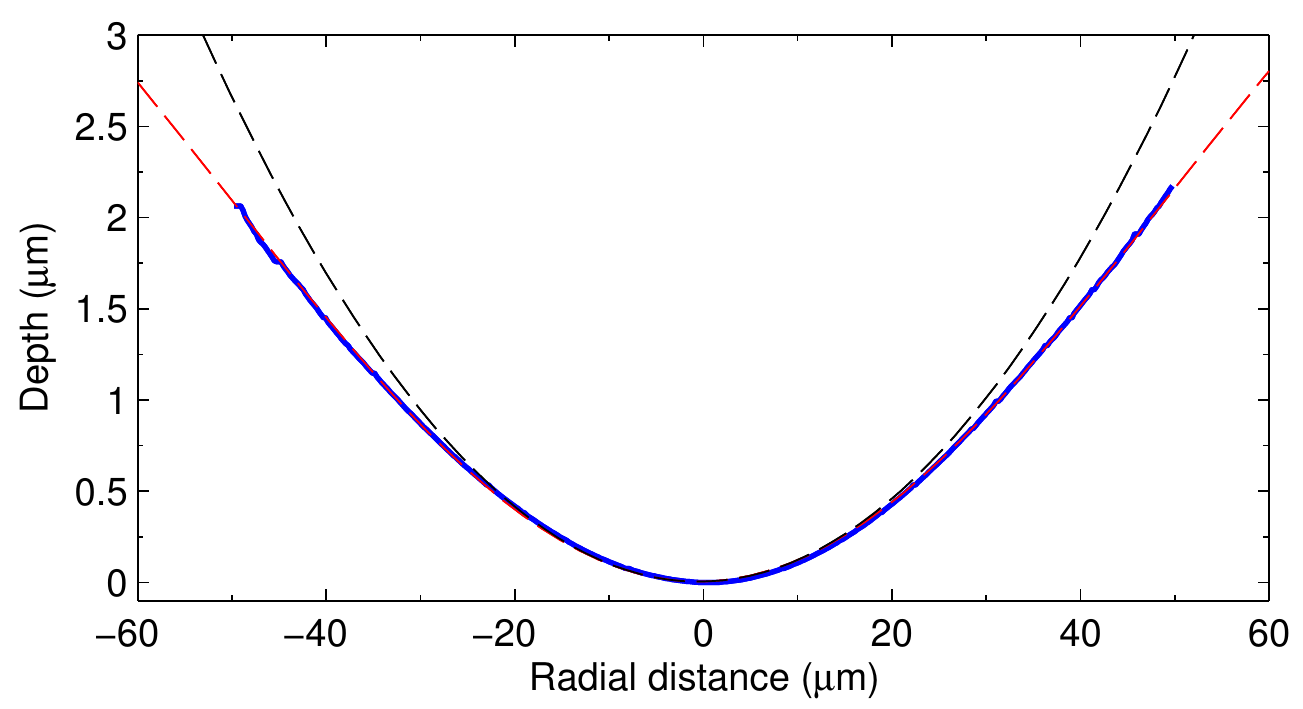}
  \label{fig:f054-cs}
 \end{subfigure}
\caption{{\bf \subref{fig:f054-opt}} An optical image of a concave fiber facet.
 {\bf \subref{fig:f054-int}} Interference image of the same fiber using white light interferometry. Here only the red layer of the color image is extracted and presented in gray scale.
{\bf \subref{fig:f054-cs}} A reconstructed cross section of the curved fiber obtained from the interference image in \subref{fig:f054-int}. The blue curve is the experimentally reconstructed fiber surface. The red dashed curve is a fit with the Gaussian function. The black dashed curve is a circle having the same local radius of curvature (457 $\mu$m) at the center.}
\label{fig:interference-example}
\end{figure}

We use white light interferometry to measure the curvature of a machined fiber with respect to a flat reference plate. For this we employ the magnetic mount described in Section \ref{sec:optical-setup}. We place a beam splitter cube on the magnetic mount, and introduce the white light from a halogen lamp along the same direction as the laser by inserting an extra mirror on the beam path while blocking the laser (see \reffig\ref{fig:interferencesetup}). The white light produces an interference pattern visible through the microscope corresponding to the topography of the fiber surface; an example is shown in \reffig\ref{fig:f054-int}. A MATLAB program analyzes the interference fringes and reconstructs the 3-dimensional shape of the surface structure (\reffig\ref{fig:f054-cs}).

The central region of the indentation is well approximated by a Gaussian profile as shown in \reffig\ref{fig:f054-cs}. This is in contrast with the results presented in \cite{Hunger2012} where it is claimed that a global deformation leads to a non-Gaussian profile. The diameter of the depression defined by the full width at $1/\e$ of the Gaussian fit is 137$\mu$m which is 1.2 and 3 times larger than the figures reported in \cite{Brandstatter2013} and \cite{Hunger2010} respectively.
Furthermore, thanks to the rotation of the fiber during machining the surface depression shows excellent symmetry around the center. The measured ellipticity is less than 2\% in \reffig\ref{fig:interference-example}.

\section{Characterization of the FFPCs}
\label{sec:char-fiber-cavit}

After laser machining, the fiber surfaces were coated with state-of-the-art high reflective coatings by AT-Films \cite{ATFilms}. The coatings are designed to have the maximum reflectivity at 866~nm with approximately 30 ppm transmission.

\begin{figure}[h!]
 \centering
 \begin{subfigure}[t]{0.45\hsize}
  \centering
  \caption{}
  \includegraphics[width=\linewidth]{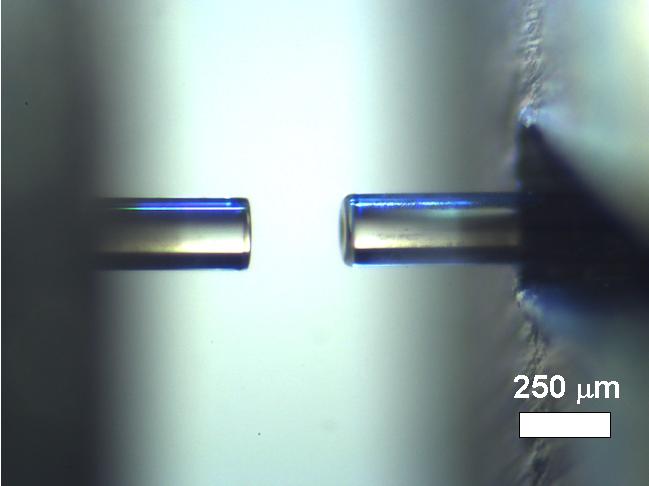}
  \label{fig:cavity_photo}
 \end{subfigure}
 \quad
 \begin{subfigure}[t]{0.45\hsize}
  \centering
  \caption{}
   \includegraphics[width=\linewidth]{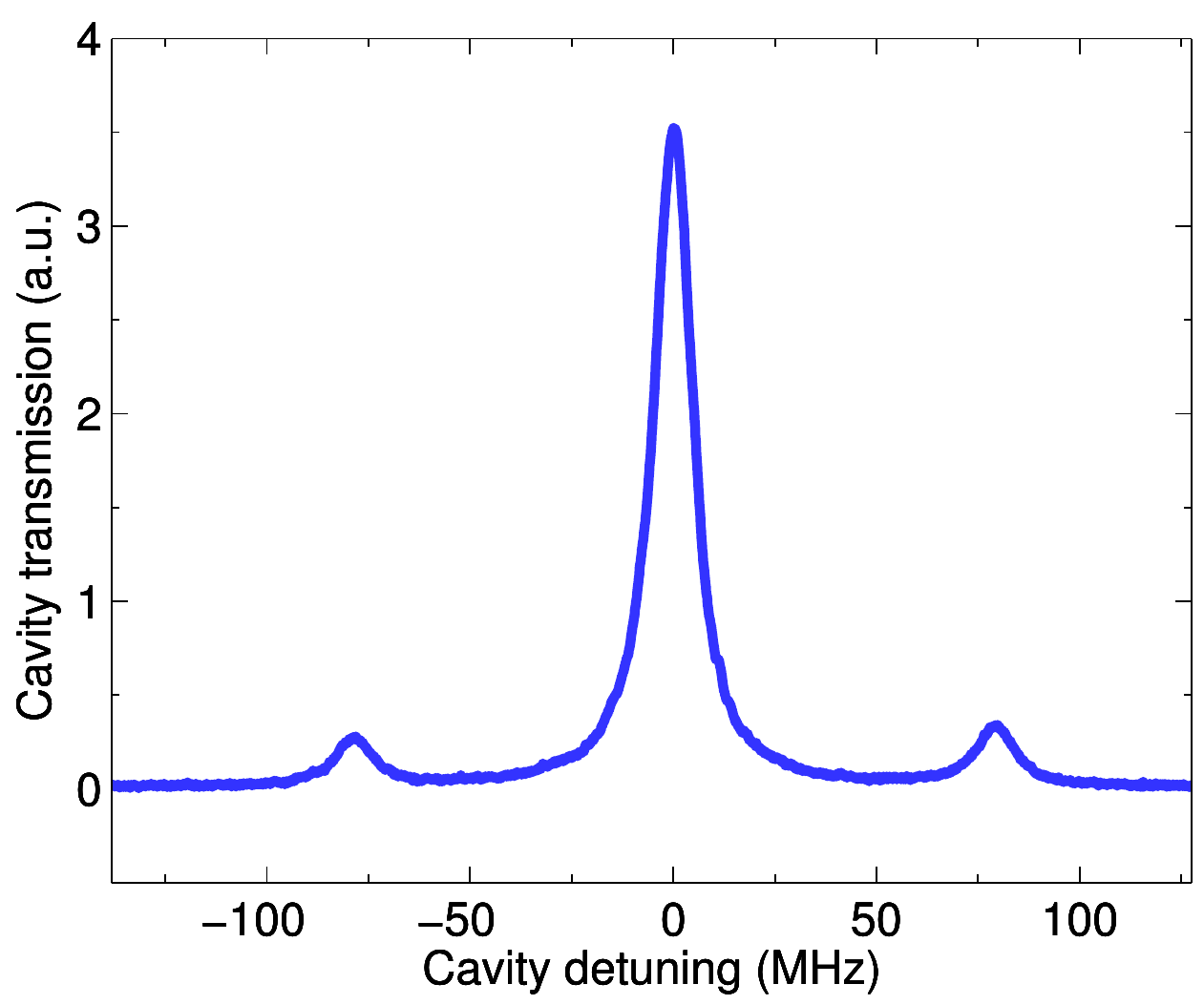}
  \label{fig:cavity_trans}
 \end{subfigure}
 \caption{{\bf \subref{fig:cavity_photo}} Microscope photo of a fiber cavity. {\bf \subref{fig:cavity_trans}} Transmission signal as the fiber cavity is scanned over resonance. The x-axis is calibrated to a detuned frequency with respect to the center of resonance by using the side band peaks at $\pm$79 MHz.}
\end{figure}

To form a Fabry-Perot cavity two fibers are mounted and aligned face-to-face as shown in \reffig\ref{fig:cavity_photo}. A laser with a wavelength of 866~nm is sent through one of the fibers and the transmission of the cavity is measured by observing the emission from the other fiber with a photo detector.
Figure \ref{fig:cavity_trans} shows the transmission signal while the cavity length is scanned over a resonance. The laser is modulated at a known radio frequency so that the transmission profile exhibits side-bands to serve as a frequency reference (\reffig\ref{fig:cavity_trans}).

The quality of an optical cavity is characterized by its finesse, given by
\begin{align}
 \mathcal{F} = \frac{c}{2L_\mr{c}\Delta},
 \label{eq:finesse}
\end{align}
where $L_\mr{c}$ is the cavity length, $\Delta$ is the full frequency width at half maximum of the cavity resonance and $c$ is the speed of light.
Another  expression of the cavity finesse can be derived in terms of the total transmissivity of the cavity mirrors $\mathcal{T}$ and sum of absorbing and scattering losses in the cavity $\mathcal{L}$:
\begin{align}
 \mathcal{F} = \frac{2\pi}{\mathcal{T}+\mathcal{L}}.
 \label{eq:finesse2}
\end{align}

The cavity length $L_\mr{c}$ is visually measured
with a calibrated microscope (see \reffig\ref{fig:cavity_photo}) and $\Delta$ is obtained by fitting a Lorentzian function to the scan profile (\reffig\ref{fig:cavity_trans}).  Using \eqref{eq:finesse}, the cavity finesse can be calculated from these two measurements.

\subsection{Impact of cavity length on finesse}
\label{sec:finesse-vs-cavity}

The dependence of the cavity finesse on the cavity length is of great interest, especially when a relatively long cavity length is required. To date, only a few measurements have been made investigating this dependence \cite{Hunger2010, Brandstatter2013}. In \cite{Hunger2010}, finesses for short cavities ($L_\mr{c}$ < 100 $\mu$m) were reported and 
observed to  drop sharply at a certain length. The authors showed that this behavior can be explained by the clipping loss model. On the other hand, in \cite{Brandstatter2013} the finesse declines gradually as the cavity length increases in contrast to the results in \cite{Hunger2010}. This behaviour is inconsistent with the clipping loss model and the cause of the drop was unknown.

We performed a series of finesse measurements with fiber pairs consisting of MM and SM fibers. We have observed two distinct behaviors depending on the types of fibers used to form the cavities. Specifically cavities formed of pairs of MM fibers behave differently from those using the combinations of SM-MM or SM-SM fibers.
\begin{figure}[htb]
 \centering
 \begin{subfigure}[c]{0.45\textwidth}
  \centering
  \caption{}
  \includegraphics[width=\linewidth]{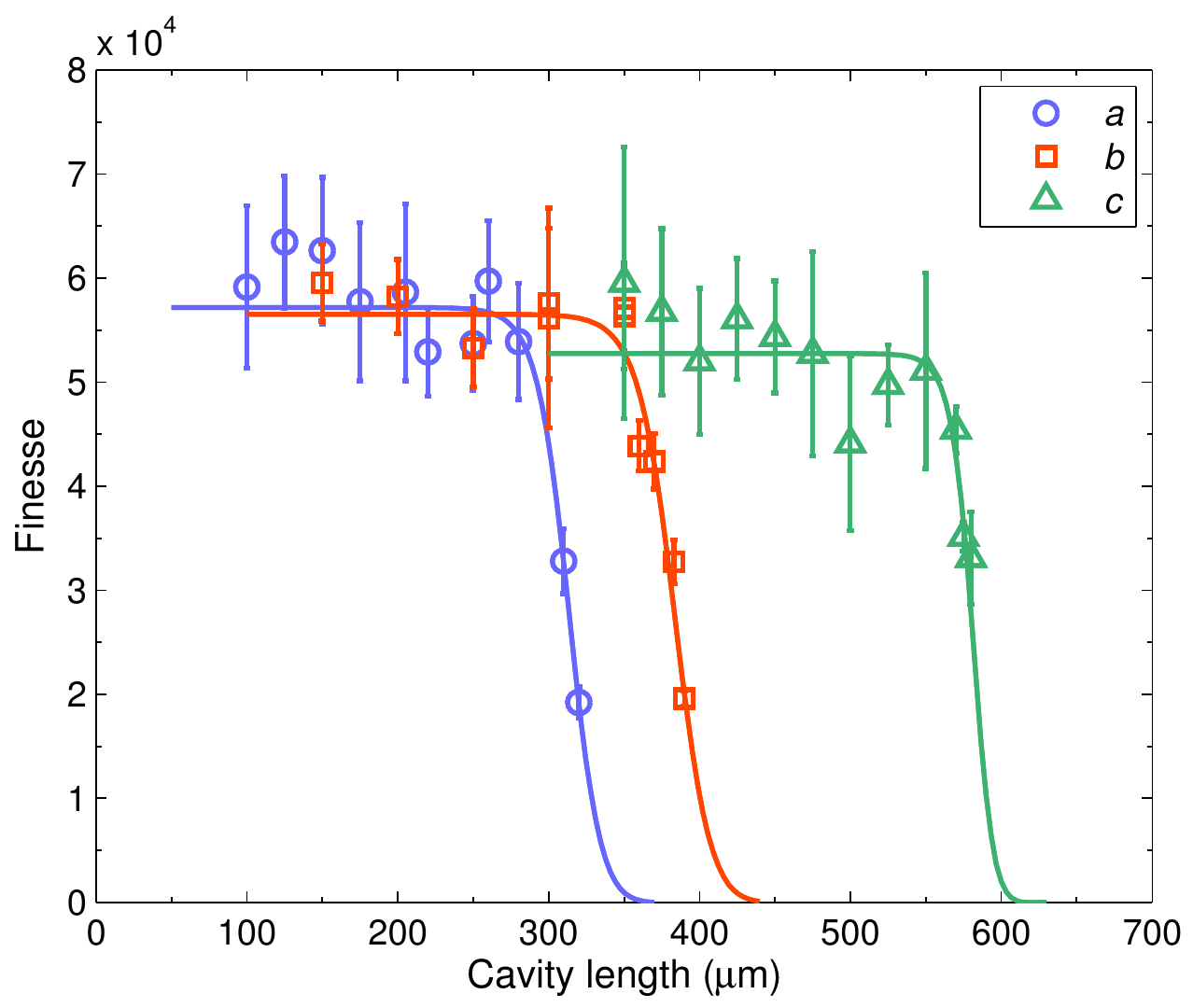}
  \label{fig:finesse-MM}
 \end{subfigure}
 \begin{subfigure}[c]{0.45\textwidth}
  \centering
  \caption{}
  \includegraphics[width=\linewidth]{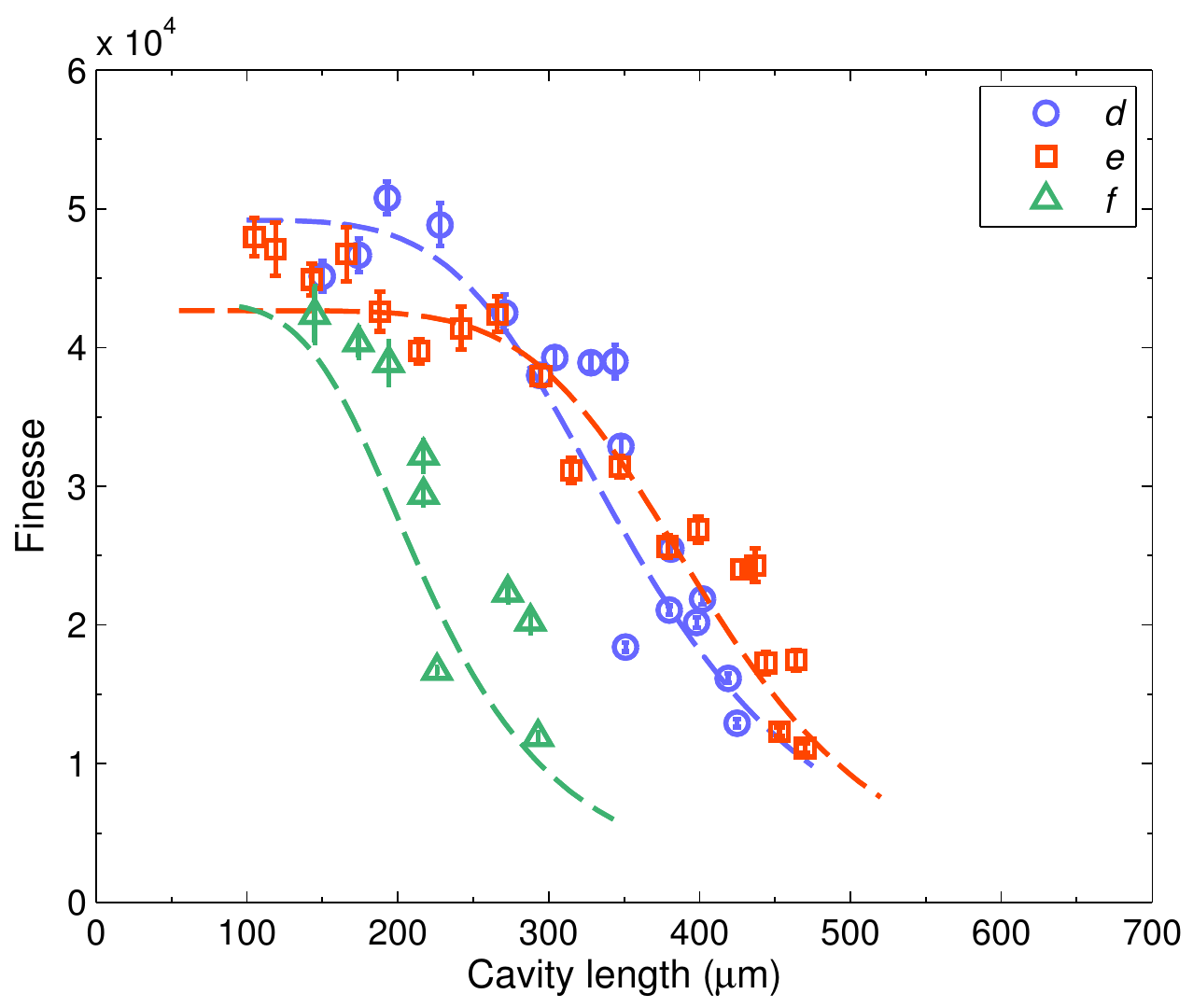}
  \label{fig:finesse-SM}
 \end{subfigure}
 \begin{subfigure}[c]{0.45\textwidth}
  \centering
  \caption{}
  \begin{tabular}[b]{| c | c c c |}\hline
      cavity & type & $R_{c1}$ ($\mu$m) & $R_{c2}$ ($\mu$m) \\ \hline
      \textit{a} & MM-MM & 469 $\pm$ 24 & $\infty$ \\
      \textit{b} & MM-MM & 669 $\pm$ 33 & $\infty$ \\
      \textit{c} & MM-MM & 669 $\pm$ 33 & 739 $\pm $37\\ \hline
      \textit{d} & MM-SM & 558 $\pm$ 28 & $\infty$ \\
      \textit{e} & MM-SM & 465 $\pm$ 24 & $\infty$ \\
      \textit{f} & SM-SM & 554 $\pm$ 28 & $\infty$ \\ \hline
  \end{tabular}
   \label{fig:cavity-table}
 \end{subfigure}
 \caption{{\bf \subref{fig:finesse-MM}} Finesse vs cavity length for 3 different MM-MM FFPCs. The error bars represent the statistical standard deviations of the measurements. The solid curves are least-square fits based on \eqref{eq:finesse-clipping} of the clipping loss model.
 {\bf \subref{fig:finesse-SM}} Finesse vs cavity length for cavities where at least one of mirrors is of SM type. The dashed curves are fits based on the clipping loss model. However here they are poorly fitted to the data.
The fitted parameters are unrealistic and have large errors. For example, $R_\mr{c}=1765\pm3645$ \micron for cavity \textit{d}. Therefore here they are shown only as eye guides.
 {\bf \subref{fig:cavity-table}} Table of cavity parameters. $R_\mr{c1}$ and $R_\mr{c2}$ are the radii of curvature of the composing mirrors at their surface centers obtained from the interferometry. $\infty$ indicates a flat surface. The measurement error is independently estimated to be 5\% by repeating interferometry with a same curved fiber surface.} 
\label{fig:finesse}
\end{figure}

Figure \ref{fig:finesse-MM} shows the finesses for the MM-MM FFPCs. The solid curves are fits based on the clipping loss model. The model assumes that upon each reflection the laser beam intensity outside some finite radius $r_{\mr{clip}}$ on the mirror surface is lost, causing a fractional loss given by
\begin{align}
 \mathcal{L}_{\mr{clip}} &= 1-\frac{\int_0^{r_{\mr{clip}}} I(r) r\,dr}{\int_0^{\infty} I(r) r\,dr} \label{eq:clipping-loss}\\
 &=\exp(-2r_{\mr{clip}}^2/w^2), \label{eq:clipping-loss2}
\end{align}
where $I(r)$ is the intensity distribution of the cavity mode on the mirror as a function of the radial distance $r$ and $w$ is the mode waist on the mirror. Note that the mode waist $w$ is a function of the radius of curvature of the cavity mirrors $R_\mr{c}$ and the cavity length $L_\mr{c}$. From \eqref{eq:clipping-loss} to \eqref{eq:clipping-loss2} we assumed that the cavity mode follows the Gaussian TEM$_{00}$ profile for spherical mirror cavities. Although the actual surface shapes of the fiber mirrors may not be spherical, it is a good approximation for small radii. In this model, using \eqref{eq:finesse2} the cavity finesse is 
\begin{align}
 \mathcal{F} &= \frac{2\pi}{\mathcal{T}+\mathcal{L}_\mr{res}+\mathcal{L}_\mr{clip}} \label{eq:finesse-clipping},
\end{align}
where $\mathcal{L}_\mr{res}$ are constant residual losses such as absorption and scattering in the coating and the clipping losses $\mathcal{L}_\mr{clip}$ are dependent on the mode diameter at the mirror surface.
All the measurements in \reffig\ref{fig:finesse-MM} are in good agreement with the fitted curves, proving the validity of the clipping loss model in this case. The large error bars of the finesse data points are mainly due to the fact that the resonance peaks are subject to fluctuations as the input light travels through a MM fiber.
From the fitted parameters all the cavities have a residual constant loss of about 35 ppm per mirror. These rather high residual losses may be significantly reduced by annealing the fibers \cite{Brandstatter2013}. 

The curve fits also provide us with estimations for $R_\mr{c}$ and $r_\mr{clip}$.
They are $R_\mr{c} = 369\pm10$ \micron and $r_\mr{clip} = 32.8\pm1.2$ \micron for cavity \textit{a} and $R_\mr{c} = 446\pm11$ \micron and $r_\mr{clip} = 36.8\pm1.3$ \micron for cavity \textit{b}.
These estimations for $R_\mr{c}$ values differ from the interferometry results by 100-200 \micron (see \reffig\ref{fig:cavity-table}). The discrepancy gets even larger for cavity \textit{c} where the fit gives $R_\mr{c} = 308\pm6.1$ \micron when the cavity is modeled as being symmetrical with curved mirrors on both sides \cite{note1}. 
These may be partially explained by improving our simple clipping loss model to explicitly incorporate diffraction losses adapted to the detailed geometry of the cavity mirrors \cite{Kleckner2010}.
Another possible source of the discrepancies is a misalignment of the fibers \cite{hauck1980misalignment}. Finesse measurements are very sensitive to the relative orientations of the two fibers, particularly when the cavity length gets long, and we note that such misalignments of the fiber axes could cause an additional diffraction loss to reduce the finesse.

In \reffig\ref{fig:finesse-SM} finesses for cavities including at least one SM fiber are shown. Here the finesses exhibit gradual decays rather than sharp drops as in \reffig\ref{fig:finesse-MM}. The decays start sooner than those in \reffig\ref{fig:finesse-MM} even though the cavity parameters are similar for the two cases. The trend suggests this is due to the introduction of the SM fibers. Furthermore, the data does not fit well with the clipping loss model (See caption of \reffig\ref{fig:finesse}).

\subsection{Formation of a ridge-like structure in machining of SM fibers}
\label{sec:formation-ridge-like}
\todo[noline]{I kept this section separate from the last one as it starts with general remarks about the SM fibre shooting and they do not very well fit in the ``Impact of cavity length on finesse'' section.}

\begin{figure}[tb]
 \centering
 \begin{subfigure}[t]{0.28\textwidth}
 \caption{} \includegraphics[width=\textwidth]{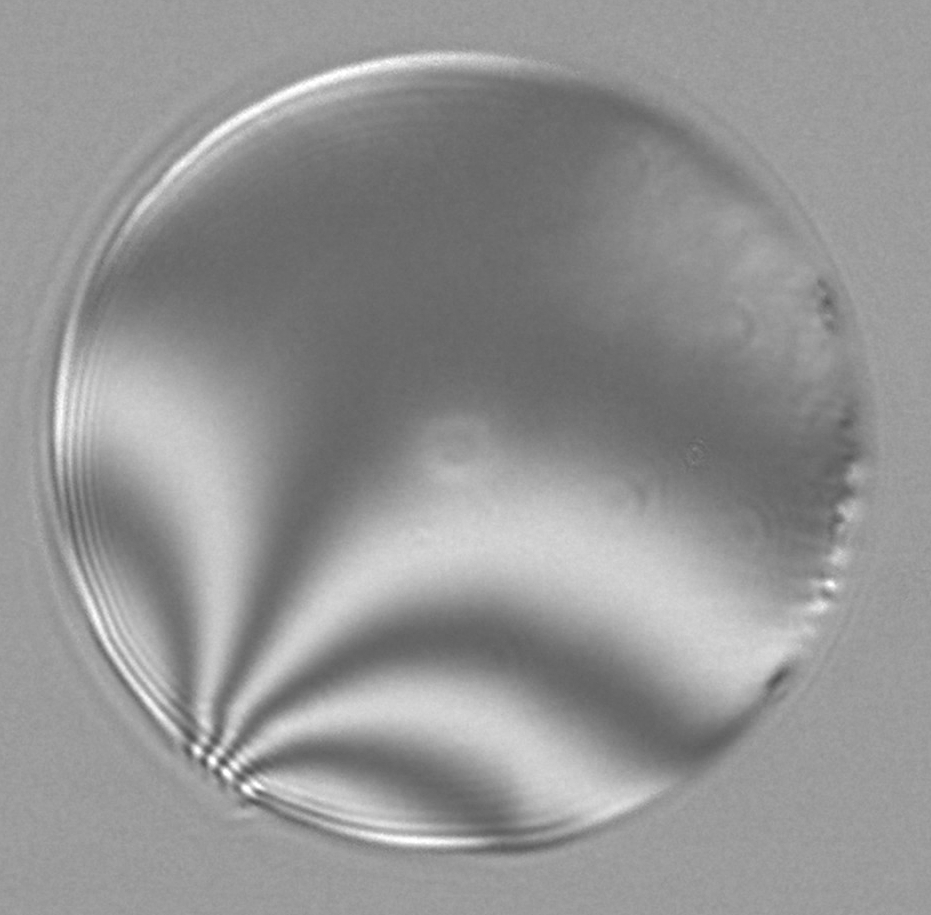}
  \label{fig:sm-core-optical}
 \end{subfigure}
 \begin{subfigure}[t]{0.28\textwidth}
  \caption{}
  \includegraphics[width=\textwidth]{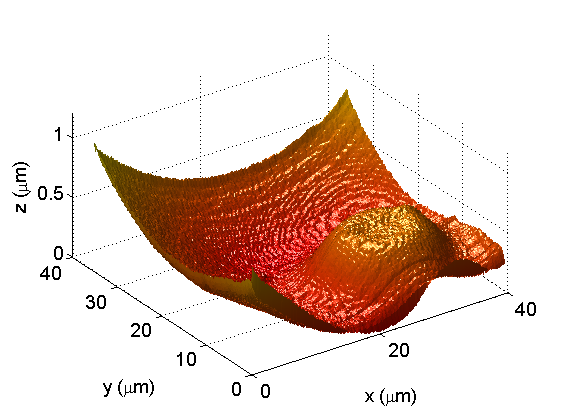}
  \label{fig:AFM-with-core}
 \end{subfigure}
 \begin{subfigure}[t]{0.28\textwidth}
  \caption{}
  \includegraphics[width=\textwidth]{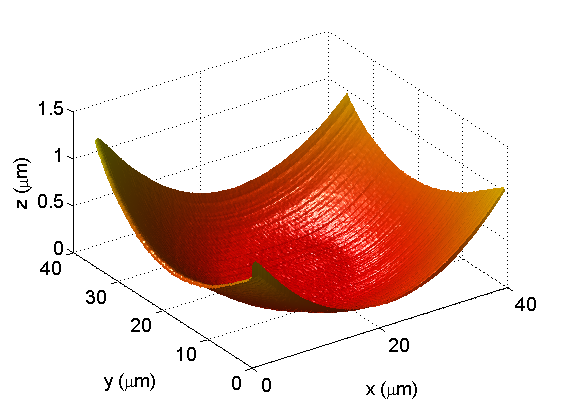}
  \label{fig:AFM-no-core}
 \end{subfigure}
 \hskip -0.4cm
 \begin{subfigure}[t]{0.1\textwidth}
  \vskip 0.2cm
  \includegraphics[width=\textwidth]{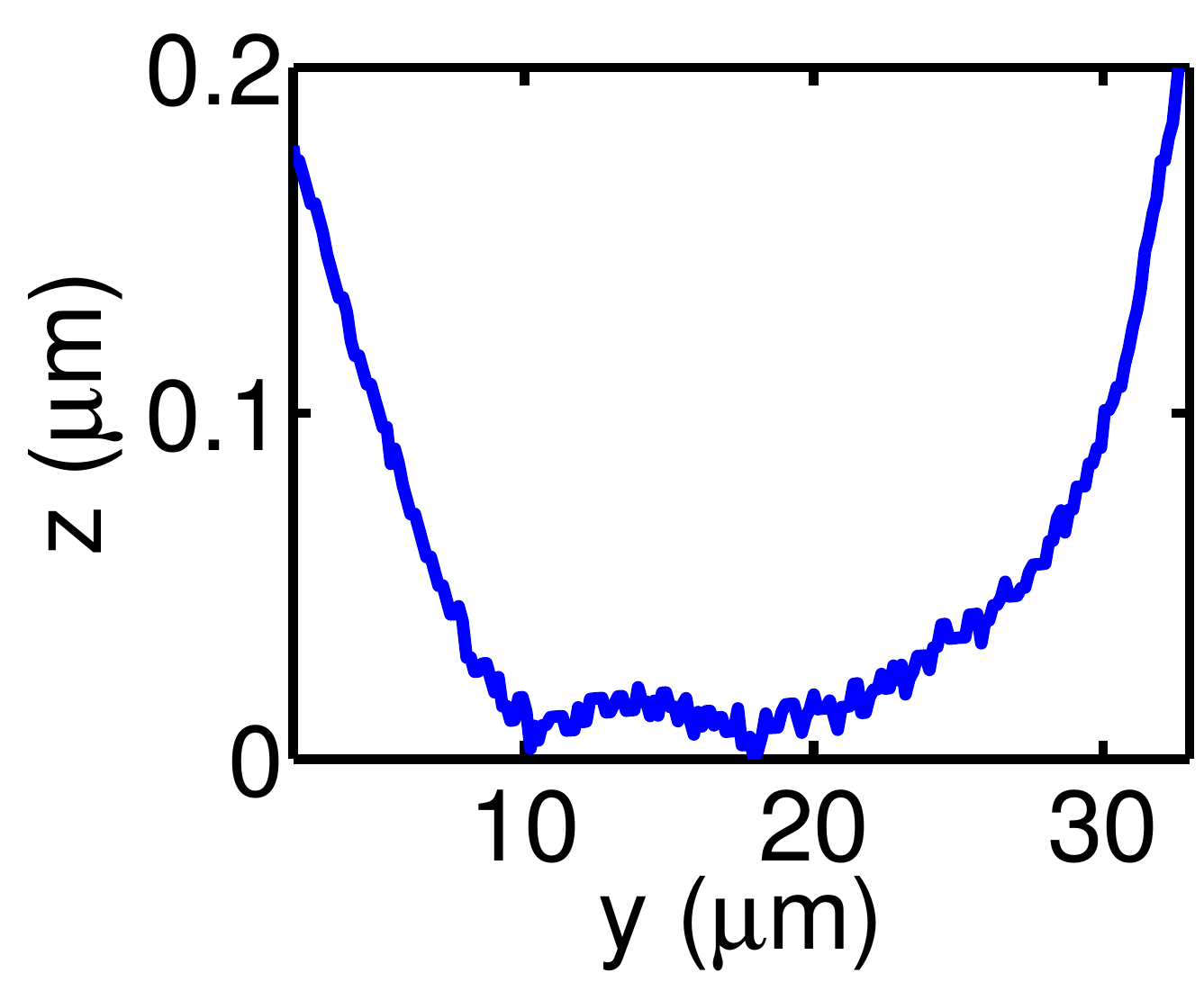}
  \label{fig:AFM-no-core-cs}
 \end{subfigure}
\caption{{\bf \subref{fig:sm-core-optical}} Interferogram of a single mode fiber. A protruding core ridge at the center is visible after the first few laser pulses. The pattern spreading from the lower left edge is due to a small deformation caused by the cleaving process. {\bf \subref{fig:AFM-with-core}} AFM image near the center of the fiber. The ridge is visible on the right. {\bf \subref{fig:AFM-no-core}} AFM image of the fiber after an attempt to remove the core ridge with further laser pulses. Here the ridge is almost totally removed. Shown in the upper right is a vertical cross section along the y-axis at x = 20 \micron. At the bottom a slightly raised section can be seen, indicating a residual of the ridge.}
\label{fig:sm_core}
\end{figure}

The MM fibers have a large diameter core which occupies almost the entire mirror surface, and are ablated uniformly during machining. However the SM fibers have a small diameter core, and most of the mirror surface is formed by the fiber cladding.
The interface between cladding and core in a SM fiber leads to a ridge-like formation when ablated (See \reffig\ref{fig:sm-core-optical}). This may be due to the difference in the thermal properties between the core and cladding materials. Normally multiple long duration laser pulses can remove the core as far as the interferogram can resolve. However this will also produce some additional curvature of the surface which is not always desirable, especially when a flat surface is intended.
We found specific machining parameters for core removal while maintaining a flat surface. To remove the core, we focused the beam to a spot with waist of $137$ \micron and used 3-5 pulses with an average pulse length of 56 ms at 2.0 W.

In order to further confirm the removal of the ridge, we made use of a Burleigh Aris 300 personal Atomic Force Microscope (AFM). This process is much slower than the optical interference method, but offers a more precise picture of the surface; the optical measurement is limited by a fraction of the wavelength of the light, while the AFM depth resolution is limited only by technical noise and our model can resolve depths of 5 nm. A ring-like ridge is clearly visible around the core of the SM fiber in \reffig\ref{fig:AFM-with-core}. On the other hand in \reffig\ref{fig:AFM-no-core} it can be seen that the ridge is largely removed after applying further laser pulses.  However a closer look reveals a possible residual of the ridge in the shape of the bottom surface as shown in the inset cross section in \reffig\ref{fig:AFM-no-core}.

In this way, although we removed the ridges to the level where they are invisible with the microscope, there are still small residual deformations which may cause measurable degradations of the cavity finesse.
For instance, a raised section seen in \reffig\ref{fig:AFM-no-core} may contribute to a defocusing of the cavity mode as the mode adapts the surface profile and hence to an additional diffraction loss. It could be also the case that a small structure from the ridge which is not fully smoothed by the laser machining but cannot be resolved by the AFM is causing a scattering loss. Even roughness in the sub-nm scale can produce a significant finesse drop in high-finesse cavities \cite{bennett1992recent}.
Therefore we suspect that the strong decays in the finesses of the SM-fiber cavities are caused by the remaining ridges created in the machining process. This is supported by the observation that cavity \textit{f} in \reffig\ref{fig:finesse-SM}, which has SM fibers for both mirrors, most rapidly decreases in finesse.

A straight-forward solution for this issue might be splicing a very short piece of MM-fiber on top of a SM-fiber. Since MM-fibers have a large core diameter ($\sim$200 \micron), a ridge formation, if any, only happens outside the region of interest around the center. 
At the same time, by keeping the section of MM-fiber short enough, the presence of the MM-fiber section would have a negligible impact on the coupling between the cavity mode and the fiber-guided mode. Besides, unwanted reflections at the splicing interface and a cavity effect inside the MM-fiber section can be avoided by matching the refractive indices of the SM and MM fibers. 

\subsection{Birefringence}
\label{sec:birefringence}

In some cQED applications where the photon polarization degree of freedom serves as a qubit, it is highly desirable to have a degenerate resonance frequency for two orthogonal polarization modes in the cavity \cite{Cirac1997}. This is usually satisfied when one uses an optical  cavity consisting of  macroscopic mirrors \cite{Ritter2012,Stute2012}. However in FFPCs  a significant birefringence, typically leading to a splitting of resonance peaks by a few hundred MHz to a few GHz, has been reported in both references \cite{Hunger2010} and \cite{Brandstatter2013}. In a recent paper, Uphoff \etal have shown that the elliptic profiles of the mirror surfaces are actually the dominant cause of birefringence in FFPCs \cite{Uphoff2014}.  
\begin{figure}[h]
 \begin{minipage}{0.48\textwidth}
  \centering
  \begin{subfigure}[c]{\linewidth}
   \caption{} \includegraphics[width=\linewidth]{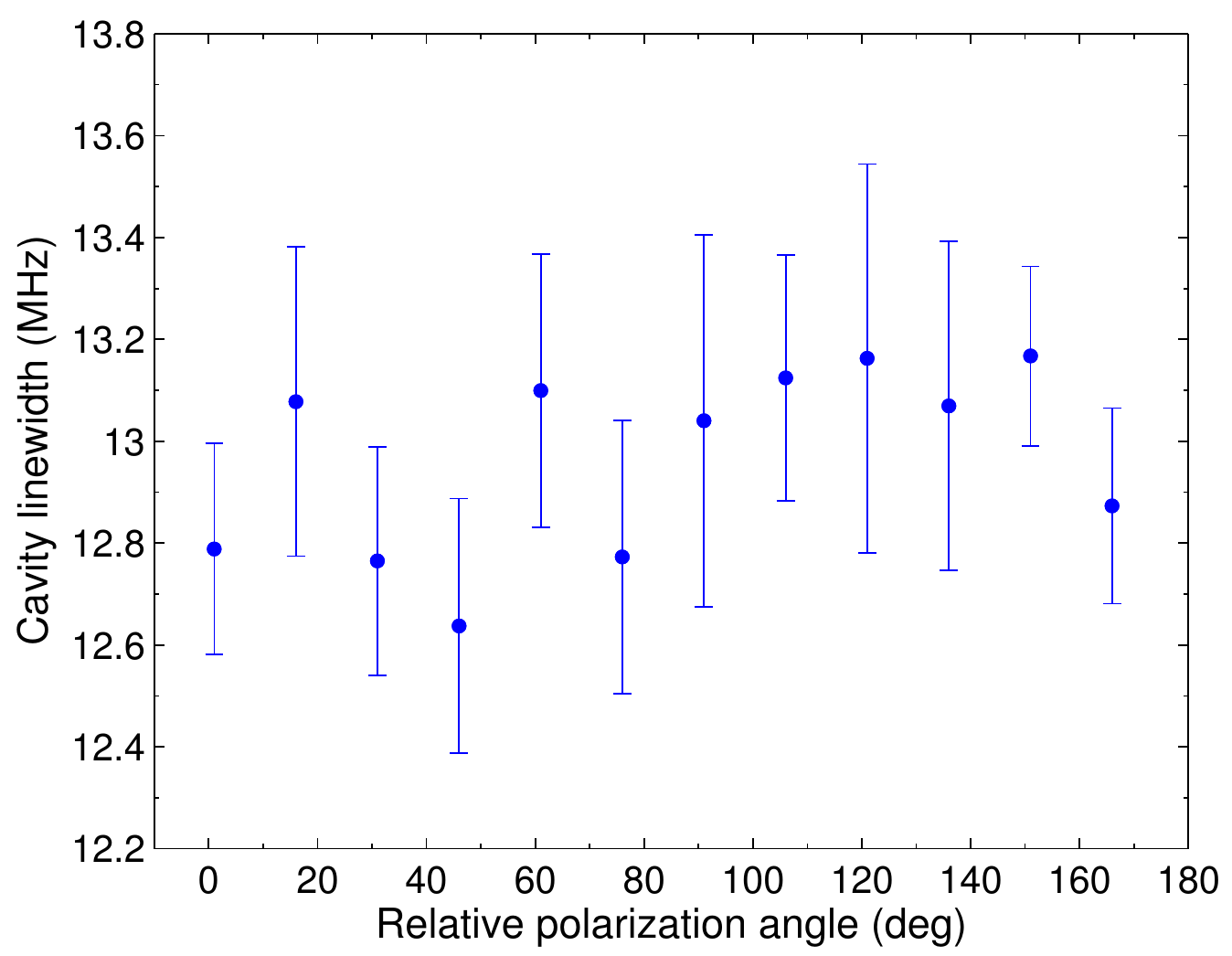}
   \label{fig:finesse-birefringence}
  \end{subfigure}
 \end{minipage}
 \hskip 10pt
 \begin{minipage}{0.48\textwidth}
  \centering
  \begin{subfigure}[c]{0.8\linewidth}
   \caption{}
  \includegraphics[width=\linewidth]{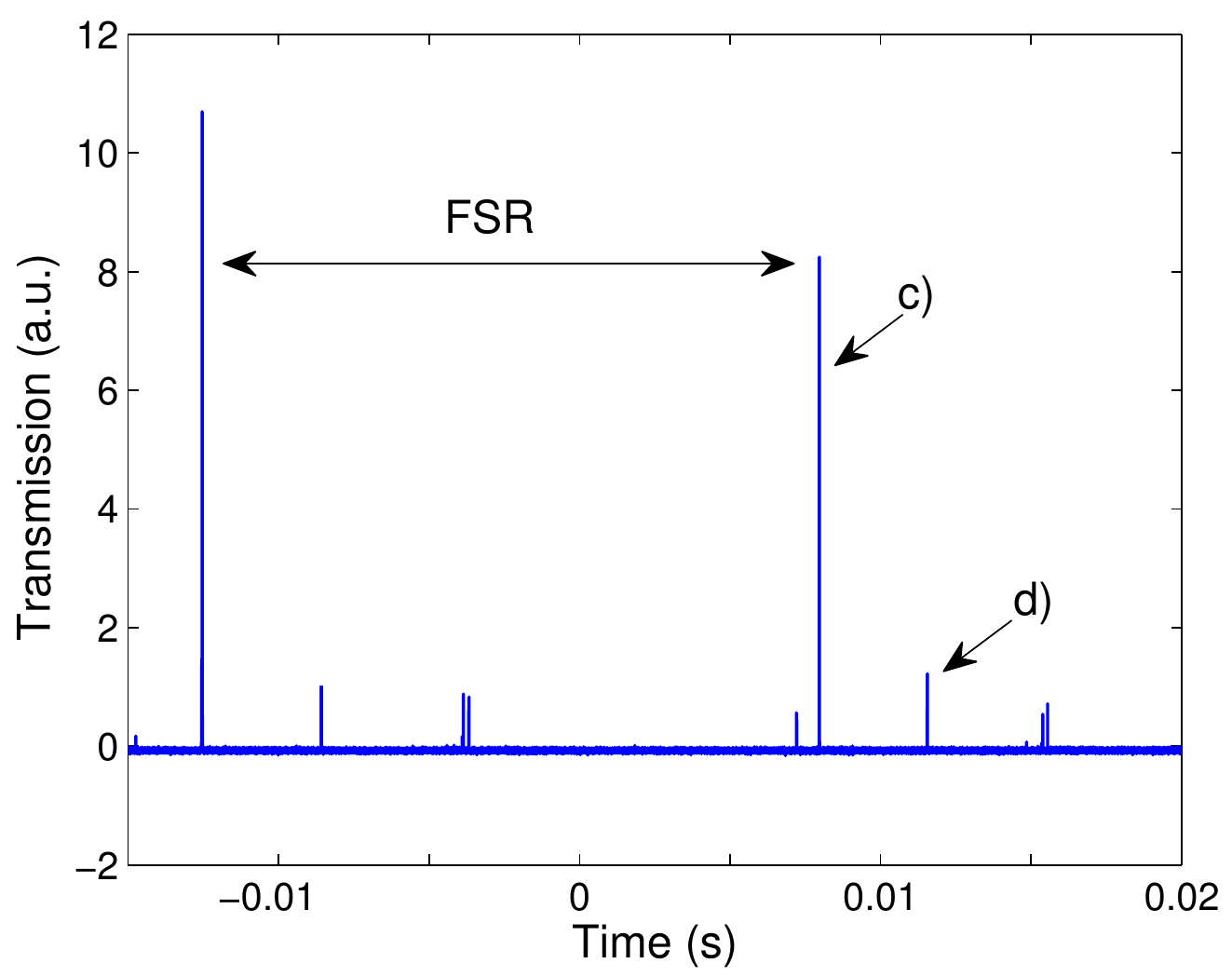}
   \label{fig:cavity-d-scan}
 \end{subfigure}
  \\
  \begin{subfigure}[t]{0.48\linewidth}
   \caption{}
   \includegraphics[width=\linewidth]{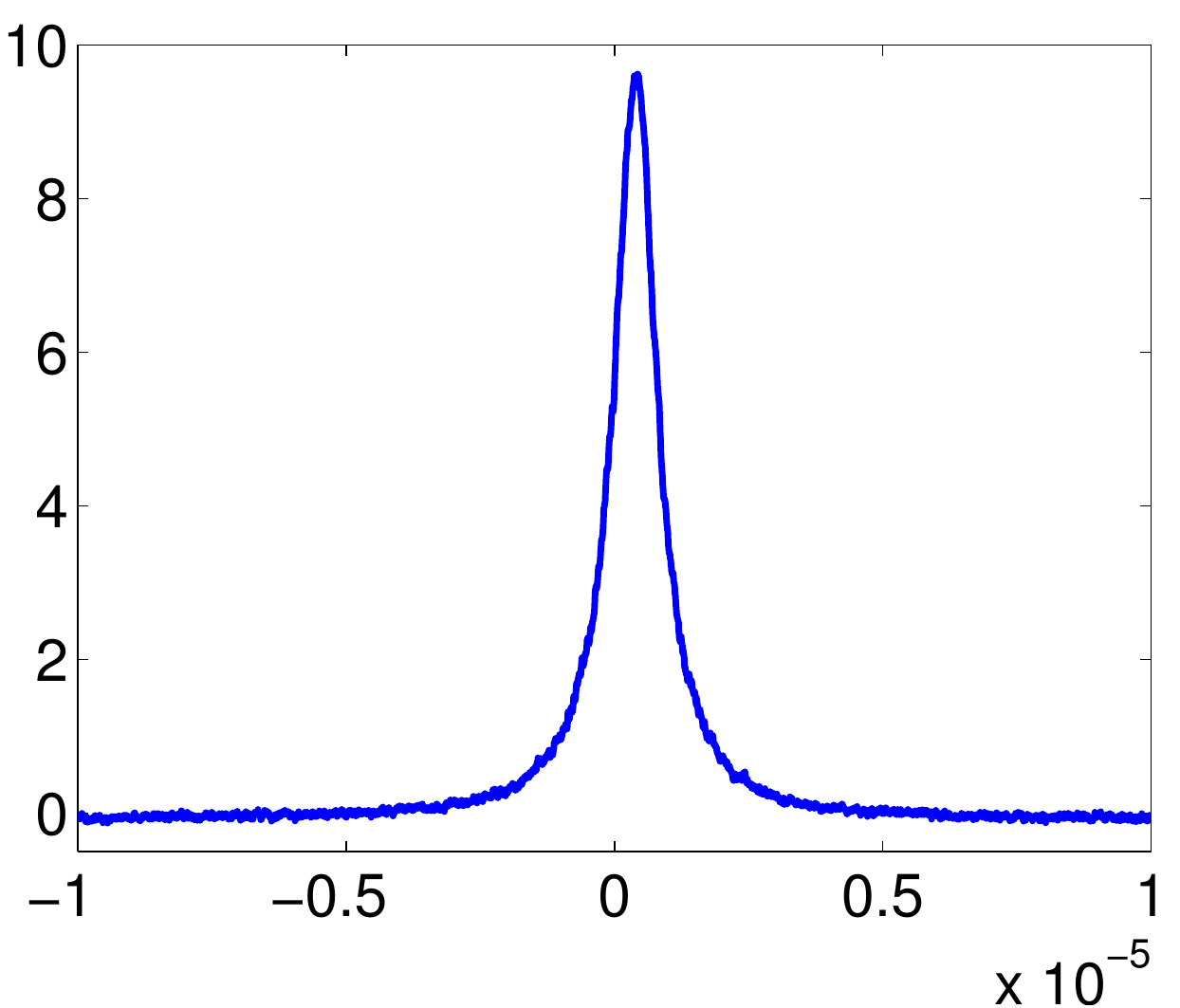}
   \label{fig:cavity-d-scan-fund}
  \end{subfigure}
  \begin{subfigure}[t]{0.48\linewidth}
   \caption{}
   \includegraphics[width=\linewidth]{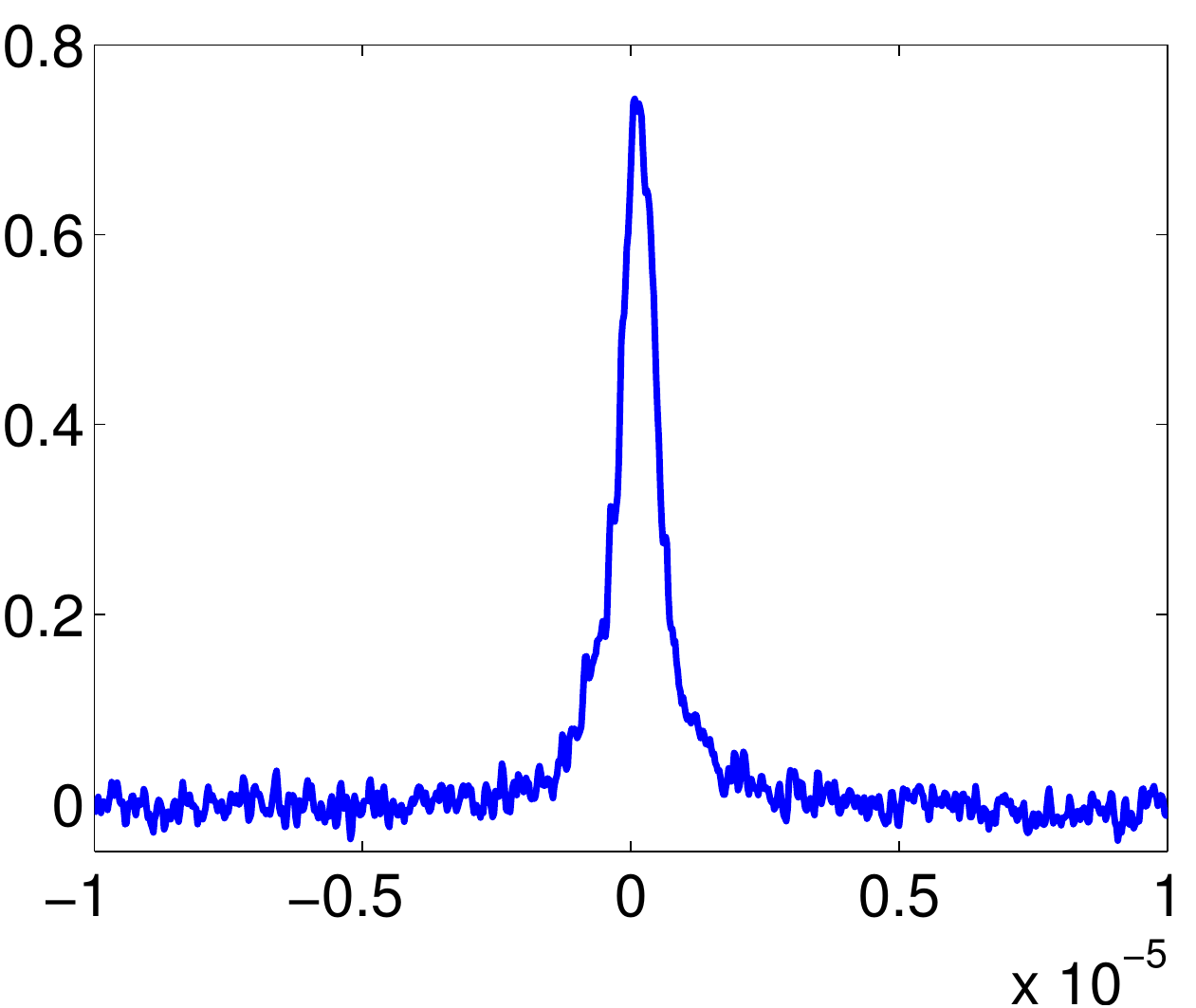}
   \label{fig:cavity-d-scan-higher-mode}
  \end{subfigure}
 \end{minipage}
 
 \caption{{\bf \subref{fig:finesse-birefringence}} Polarization dependence of the linewidth (FWHM) of the fundamental mode in cavity \textit{d}. The error bars represent one statistical standard deviation. %
 {\bf \subref{fig:cavity-d-scan}} A typical transmission profile of cavity \textit{d} when its length is scanned over time. FSR indicates a change of the cavity length corresponding to one  free spectral range. %
 {\bf \subref{fig:cavity-d-scan-fund}}, {\bf \subref{fig:cavity-d-scan-higher-mode}} Close-up views of the scan profile for the fundamental and the next higher mode respectively as indicated in \subref{fig:cavity-d-scan}.}
 \label{fig:birefringence}
\end{figure}

We measured the dependence of a cavity linewidth on the input polarization using the fundamental transverse mode of cavity \textit{d} used in \reffig\ref{fig:finesse}. The cavity length was set to 270 \micron. Since the TEM modes inside the cavity are not preserved in the output fiber, we cannot explicitly examine the field distributions of different transverse modes in the cavity. However due to the configuration of cavity \textit{d} to have a SM-fiber with a flat surface mirror on the input side, the mode matching between the input mode and the cavity's TEM$_{00}$ mode is expected to be dominant. Hence it is reasonable to assume that the largest peak in the scan in \reffig\ref{fig:cavity-d-scan} corresponds to the fundamental TEM$_{00}$ mode. The input light is linearly polarized and rotated with a half wave plate before being coupled into the input SM fiber.
The cavity linewidth shows little dependence on the polarization of the input light and measured values coincide within the error bars, as shown in \reffig\ref{fig:finesse-birefringence}.
Also there is no visible splitting of the resonance peaks, which would otherwise manifest birefringence, in the cavity's fundamental mode as well as the higher order modes (see \reffig\ref{fig:cavity-d-scan-fund} and \reffig\ref{fig:cavity-d-scan-higher-mode}).

To obtain an upper limit for the variation of the cavity linewidth due to the input polarization, we take the difference between the highest and lowest points within the error bars in \reffig\ref{fig:finesse-birefringence}. This corresponds to a change of the linewidth of 1.16 MHz, with the mean value being 13.0 MHz.
In order to reproduce this change of the linewidth solely by a splitting of the polarization modes, the splitting in frequency needs to be 3.84 MHz. This was determined by fitting a single Lorentzian function to a simulated  double-peaked Lorentzian profile with a given peak separation. Note that this value is an overestimation of the birefringence as the errors in \reffig\ref{fig:finesse-birefringence} also include the technical and statistical noise in the measurements.
In \cite{Uphoff2014} the frequency splitting of the orthogonal polarization modes in an elliptic mirror cavity is given by
\begin{align}
 \Delta\nu =\frac{\lambda\nu_{\mathrm{FSR}}}{(2\pi)^2}\frac{R_1-R_2}{R_1R_2},
\end{align}
where $\nu_{\mathrm{FSR}}$ is the free spectral range in frequency and $R_1$ and $R_2$ are the radii of curvature along the principal axes of the elliptic paraboloidal shape of the surface. Using the mirror parameters of cavity \textit{d} we get $\Delta\nu = 2.36$ MHz which is close to the aforementioned value and well below the cavity linewidth.

These results distinguish themselves from those in \cite{Hunger2010, Brandstatter2013} and demonstrate that our FFPCs are readily usable for many applications of cQED where coherent mapping of light polarizations into and out of an atomic degree of freedom is required \cite{Cirac1997, Stute2012}.   

\section{Conclusion}
\label{sec:conclusion}

We have established a novel method to laser-machine the surfaces of optical fiber facets using multiple laser pulses and fiber rotation. This method allows for the production of a highly symmetric and uniform structure on a fiber surface across a region larger than 100 \micron in diameter.
 
We characterized the length and polarization dependence of the cavity finesse for these FFPCs.
We found that the cavities formed by the MM fibers showed a constant cavity finesse over a wide range of cavity lengths of up to several hundreds of \micron and are qualitatively consistent with the clipping loss model. On the other hand the cavities using a SM fiber exhibited a gradual decline in finesse starting at a shorter cavity length. This premature decay might originate from a small ridge structure around the fiber core created during the machining process. Nevertheless their finesses were stable up to a cavity length of 200 $\mu$m. Furthermore, our FFPCs showed an excellent polarization characteristic, with no measurable birefringence.

These results demonstrate that our FFPCs can be readily utilized in future cQED experiments requiring micro-resonators. 
Our birefringence-free FFPCs are suitable over a wide range of cavity lengths, which benefit long cavity requirement applications such as ion-trap cQED, and when degenerate polarization modes are essential.

\section*{Acknowledgments}
We gratefully acknowledge support from Japan Science and Technology Agency (PRESTO) and the EPSRC (EP/J003670/1).


\begin{thebibliography}{10}
\newcommand{\enquote}[1]{``#1''}

\bibitem{Kimble2008}
H.~J. Kimble, \enquote{{The quantum internet,}} Nature \textbf{453}, 1023--1030
  (2008).

\bibitem{Vernooy1998a}
D.~W. Vernooy, V.~S. Ilchenko, H.~Mabuchi, E.~W. Streed, and H.~J. Kimble,
  \enquote{{High-Q measurements of fused-silica microspheres in the near
  infrared,}} Opt. Lett. \textbf{23}, 247--249 (1998).

\bibitem{Armani2003}
D.~Armani, T.~Kippenberg, S.~Spillane, and K.~Vahala, \enquote{{Ultra-high-Q
  toroid microcavity on a chip},} Nature \textbf{421}, 925--928 (2003).

\bibitem{Hunger2010}
D.~Hunger, T.~Steinmetz, Y.~Colombe, C.~Deutsch, T.~W. H\"{a}nsch, and
  J.~Reichel, \enquote{{A fiber Fabry-Perot cavity with high finesse},} New
  J. Phys. \textbf{12}, 065038 (2010).

\bibitem{pollinger2009ultrahigh}
M.~P{\"o}llinger, D.~O'Shea, F.~Warken, and A.~Rauschenbeutel,
  \enquote{{Ultrahigh-Q tunable whispering-gallery-mode microresonator},}
  Phys. Rev. Lett. \textbf{103}, 053901 (2009).

\bibitem{Thompson2013}
J.~Thompson, T.~Tiecke, N.~de~Leon, J.~Feist, A.~Akimov, M.~Gullans, A.~Zibrov,
  V.~Vuleti{\'c}, and M.~Lukin, \enquote{{Coupling a Single Trapped Atom to a
  Nanoscale Optical Cavity},} Science \textbf{587}, 1202--1205 (2013).

\bibitem{Brandstatter2013}
B.~Brandst\"{a}tter, A.~McClung, K.~Sch\"{u}ppert, B.~Casabone, K.~Friebe,
  A.~Stute, P.~O. Schmidt, C.~Deutsch, J.~Reichel, R.~Blatt, and T.~E. Northup,
  \enquote{{Integrated fiber-mirror ion trap for strong ion-cavity coupling,}}
  Rev. Sci. Instrum. \textbf{84}, 123104 (2013).

\bibitem{Colombe2007}
Y.~Colombe, T.~Steinmetz, G.~Dubois, F.~Linke, D.~Hunger, and J.~Reichel,
  \enquote{{Strong atom-field coupling for Bose-Einstein condensates in an
  optical cavity on a chip},} Nature \textbf{450}, 272--276 (2007).

\bibitem{Steiner2013}
M.~Steiner, H.~M. Meyer, C.~Deutsch, J.~Reichel, and M.~K\"{o}hl,
  \enquote{{Single Ion Coupled to an Optical Fiber Cavity},} Phys. Rev.
  Lett. \textbf{110}, 043003 (2013).

\bibitem{Flowers-Jacobs2012}
N.~E. Flowers-Jacobs, S.~W. Hoch, J.~C. Sankey, A.~Kashkanova, a.~M. Jayich,
  C.~Deutsch, J.~Reichel, and J.~G.~E. Harris, \enquote{{Fiber-cavity-based
  optomechanical device},} Appl. Phys. Lett. \textbf{101}, 221109 (2012).

\bibitem{Hunger2012}
D.~Hunger, C.~Deutsch, R.~J. Barbour, R.~J. Warburton, and J.~Reichel,
  \enquote{{Laser micro-fabrication of concave, low-roughness features in
  silica},} AIP Advances \textbf{2}, 012119 (2012).

\bibitem{Hijlkema2007}
M.~Hijlkema, B.~Weber, H.~P. Specht, S.~C. Webster, A.~Kuhn, and G.~Rempe,
  \enquote{{A single-photon server with just one atom},} Nature Phys.
  \textbf{3}, 253--255 (2007).

\bibitem{Ritter2012}
S.~Ritter, C.~N\"{o}lleke, C.~Hahn, A.~Reiserer, A.~Neuzner, M.~Uphoff,
  M.~M\"{u}cke, E.~Figueroa, J.~Bochmann, and G.~Rempe, \enquote{{An elementary
  quantum network of single atoms in optical cavities},} Nature \textbf{484},
  195--200 (2012).

\bibitem{ATFilms}
Advanced Thin Films, 5733 Central Avenue Boulder, Colorado USA.

\bibitem{note1}
Cavity \textit{c} is actually asymmetric. However a model with two redundant
  parameters $R_\mr{c1}$ and $R_\mr{c2}$ converges to an indecisive fit with
  large parameter errors. Therefore here we deduce an effective radius of
  curvature using the symmetric cavity model.

\bibitem{Kleckner2010}
D.~Kleckner, W.~T.~M. Irvine, S.~S.~R. Oemrawsingh, and D.~Bouwmeester,
  \enquote{{Diffraction-limited high-finesse optical cavities},} Phys.
  Rev. A \textbf{81}, 043814 (2010).

\bibitem{hauck1980misalignment}
R.~Hauck, H.~Kortz, and H.~Weber, \enquote{Misalignment sensitivity of optical
  resonators,} Appl. Opt. \textbf{19}, 598--601 (1980).

\bibitem{bennett1992recent}
J.~M. Bennett, \enquote{Recent developments in surface roughness
  characterization,} Meas. Sci. Technol. \textbf{3}, 1119--1127
  (1992).

\bibitem{Cirac1997}
J.~I. Cirac, P.~Zoller, H.~J. Kimble, and H.~Mabuchi, \enquote{{Quantum state
  transfer and entanglement distribution among distant nodes in a quantum
  network},} Phys. Rev. Lett. \textbf{78}, 3221--3224 (1997).

\bibitem{Stute2012}
A.~Stute, B.~Casabone, P.~Schindler, T.~Monz, P.~O. Schmidt,
  B.~Brandst\"{a}tter, T.~E. Northup, and R.~Blatt, \enquote{{Tunable
  ion-photon entanglement in an optical cavity},} Nature \textbf{485}, 482--485
  (2012).

\bibitem{Uphoff2014}
M.~Uphoff, M.~Brekenfeld, G.~Rempe, and S.~Ritter, \enquote{{Frequency
  splitting of polarization eigenmodes in microscopic Fabry-Perot cavities},}
  arXiv:1408.4367  (2014).

\end{thebibliography}
\end{document}